# ROTOMAGNETIC COUPLING IN FINE-GRAINED MULTIFERROIC BiFeO₃: THEORY AND EXPERIMENT

**Anna N. Morozovska[1],[\*], Eugene A. Eliseev[2], Maya D. Glinchuk[2], Olena M. Fesenko[1], Vladimir V. Shvartsman[6], Venkatraman Gopalan[3], Maxim V. Silibin[4][†], and Dmitry V. Karpinsky[5][‡]**

[1] *Institute of Physics, National Academy of Sciences of Ukraine,*

*46, pr. Nauky, 03028 Kyiv, Ukraine*

[2] *Institute for Problems of Materials Science, National Academy of Sciences of Ukraine, Krjijanovskogo 3, 03142 Kyiv, Ukraine*

[3] *Department of Materials Science and Engineering, Pennsylvania State University, University Park, PA 16802, USA*

[4] *National Research University of Electronic Technology "MIET", Moscow, Zelenograd, Russia*

[5] *Scientific-Practical Materials Research Centre of NAS of Belarus, Minsk, Belarus*

[6]*Institute for Material Science and Center for Nanointegration Duisburg-Essen (CENIDE), University of Duisburg-Essen, Universitätsstraße 15, 45141 Essen, Germany*

## Abstract

Using Landau-Ginzburg-Devonshire (LGD) theory for BiFeO₃ dense fine-grained ceramics with quasi-spherical grains and nanosized inter-grain spaces enriched by elastic defects, we calculated a surprisingly strong size-induced increase of the AFM temperature caused by the joint action of rotomagnetic and magnetostrictive coupling. Notably that all parameters included in the LGD functional have been extracted from experiments, not assumed. Complementary we performed experiments for dense BiFeO₃ ceramics, which revealed that the shift of antiferromagnetic transition to $T_N$ ~690 K instead of $T_N$~645 K for a single crystal. To explain theoretically the result, we consider the possibility to control antiferromagnetic state of multiferroic BiFeO₃ via biquadratic antiferrodistortive roto-magnetic, roto-electric, magnetostrictive and magnetoelectric couplings. According to our calculations the highest is the rotostriction contribution, the magnetostrictive and electrostriction contributions appeared smaller.

[\*] Corresponding author: anna.n.morozovska@gmail.com
[†] Corresponding author: sil_m@mail.ru
[‡] Corresponding author: dmitry.karpinsky@gmail.com

# I. INTRODUCTION

Multiferroics, defined as materials with more than one interacting long-range ferroic orders, are ideal systems for fundamental studies of couplings among the order parameters of different nature: e.g. ferroelectric polarization, structural antiferrodistortion, and antiferromagnetic order parameters [1, 2, 3, 4, 5, 6, 7, 8].

The antiferrodistortive (**AFD**), polar and magnetic degrees of freedom in multiferroics are linked via the different types of biquadratic couplings leading to versatile phase diagrams and domain structure evolution [1-8]. The biquadratic couplings are universal for the AFD multiferroics [9]. In particular, the "rotoelectric" Houchmandazeh-Laizerowicz-Salje coupling is the biquadratic coupling between the AFD order parameter and polarization [10, 11, 12]. The "direct" rotomagnetic coupling is the biquadratic coupling between the AFD and (anti)ferromagnetic orders [13].

Among the couplings, the rotomagnetic coupling impact is the most poorly studied experimentally and theoretically, except experiments by Bussmann-Holder et al. [14, 15] revealing a magnetic field impact on AFD tilts in EuTiO$_3$. The goal of the present work is to study theoretically the impact of rotomagnetic and rotoelectric couplings on antiferromagnetic (**AFM**) order of multiferroic bismuth ferrite BiFeO$_3$ (**BFO**).

BFO is the one of the rare multiferroics with antiferromagnetism and a strong ferroelectric polarization at room temperature, as well as conduction and magnetotransport on domain walls [16, 17, 18, 19]. The pronounced multiferroic properties and unique and domain structure evolution maintain in BFO thin films and heterostructures [20, 21, 22, 23, 24, 25]. Bulk BFO exhibits antiferrodistortive (AFD) order at temperatures below 1200 K; it is ferroelectric (FE) with a large spontaneous polarization below 1100 K and is AFM below Neel temperature $T_N \approx 650$ K [26, 8]. Recently full phase diagram of BFO including AFM, FE, and AFD phases was calculated within Landau-Ginzburg-Devonshire (**LGD**) theory [27], however the role of rotomagnetic and rotoelectric couplings was omitted.

This work using LGD-theory establish the rotomagnetic coupling influence on the AFM transition temperature of BFO ceramics with quasi-spherical micron sized grains (treated as a core) and nanosized inter-grain spaces (shell). Notably that all parameters included in the LGD functional have been extracted from experiments, not assumed. Then we will present experimental results for dense BFO ceramics, which reveal shift of the AFM transition to $T_N \sim 690$ K instead of $T_N \sim 645$ K for a single crystal, and performed theoretical estimates of possible contributions to the shift. According to the estimates the highest is the rotomagnetic coupling contributions, the magnetoelectric and rotoelectric ones are smaller and much smaller, respectively.

The original part of the paper is organized as follows. The impact of rotomagnetic, rotoelectric and magnetoelectric couplings on the AFM transition of BFO is presented in Section II. Experimental results are analyzed Section III. Section IV is devoted to the discussion of the agreement between theory and experiment. Section V is a conclusions.

## II. THEORETICAL DESCRIPTION

Rotomagnetic coupling is described by the term $\left(\xi_{ijkl}^M M_i M_j + \xi_{ijkl}^L L_i L_j\right) \Phi_k \Phi_l$, where $L_i$ is antiferromagentic order parameter. Rotoelectric coupling is described by the term $\xi_{ijkl} P_i P_j \Phi_k \Phi_l$, where $\xi_{ijkl}$ is the rotoelectric coupling coefficient, $P_i$ is the spontaneous polarization vector, $\Phi_i$ are the spontaneous oxygen octahedra tilt angles [9]. Biquadratic magnetoelectric coupling is the coupling between polarization and magnetization, that is described by the term $\eta_{ijkl} P_i P_j M_k M_l$, where $\eta_{ijkl}$ is the biquadratic magnetoelectric coupling coefficient and $M_i$ is the spontaneous magnetization.

Thermodynamic potential of Landau-Ginzburg-Devonshire (**LGD**) type that describes antiferromagnetic (**AFM**), ferroelectric (**FE**) and antiferrodistortive (**AFD**) properties of $BiFeO_3$, including the rotomagnetic, rotoelectric and magnetoelectric biquadratic couplings includes the AFD, FE, AFM contributions and the coupling ($\Delta G_{BQC}$) among them [27], as well elastic energy ($\Delta G_{ELS}$) including electrostrictive, magnetostrictive, and rotostrictive contributions existing in a strained media:

$$\Delta G = \Delta G_{AFD} + \Delta G_{FE} + \Delta G_{AFM} + \Delta G_{BQC} + \Delta G_{ELS} \tag{1}$$

Below we are mainly interested in $R3c$ phase that has nonzero $\Phi = \Phi_1 = \Phi_2 = \Phi_3$ and $P = P_1 = P_2 = P_3$ and G-type (cycloidal) dimensionless AFM order parameter, $L = \left(M_a - M_b\right)/2M_0$, existing below Neel temperature. The AFD energy in the $R3c$ phase is a six-order expansion on the oxygen tilt $\Phi_i$ and its gradients,

$$\Delta G_{AFD} = a_i^{(\Phi)} \Phi_i^2 + a_{ij}^{(\Phi)} \Phi_i^2 \Phi_j^2 + a_{ijk}^{(\Phi)} \Phi_i^2 \Phi_j^2 \Phi_k^2 + g_{ijkl}^{(\Phi)} \frac{\partial \Phi_i}{\partial x_k} \frac{\partial \Phi_j}{\partial x_l} \tag{2}$$

Here $\Phi_i$ are components of pseudovectors, determining out-of-phase static rotations of oxygen octahedral groups (eigenvectors of AFD modes of lattice vibrations), and Einstein summation convention is employed.

FE energy $\Delta G_{FE}$ is a six-order expansion on the polarization vector $P_i$ and its gradients,

$$\Delta G_{FE} = a_i^{(P)} P_i^2 + a_{ij}^{(P)} P_i^2 P_j^2 + a_{ijk}^{(P)} P_i^2 P_j^2 P_k^2 + g_{ijkl}^{(P)} \frac{\partial P_i}{\partial x_k} \frac{\partial P_j}{\partial x_l}. \qquad (3)$$

AFM energy $\Delta G_{AFM}$ is a fourth-order expansion in terms of the AFM order parameter vector $L_i$ and its gradient because this phase transition in BiFeO$_3$ is known to be the second order one.

$$\Delta G_{AFM} = a_i^{(L)} L_i^2 + a_{ij}^{(L)} L_i^2 L_j^2 + g_{ijkl}^{(L)} \frac{\partial L_i}{\partial x_k} \frac{\partial L_j}{\partial x_l}. \qquad (4)$$

In accordance with the classical LGD theory, we assume that the coefficients $a_i^{(\Phi)}$ and $a_k^{(P)}$ are temperature dependent according to Barrett law [28], $a_i^{(\Phi)} = \alpha_T^{(\Phi)} T_{q\Phi} \left( \coth\left( T_{q\Phi}/T \right) - \coth\left( T_{q\Phi}/T_{\Phi} \right) \right)$ and $a_k^{(P)} = \alpha_T^{(P)} \left( T_{qP} \coth\left( T_{qP}/T \right) - T_C \right)$, where $T_{\Phi}$ and $T_C$ are corresponding virtual Curie temperatures, $T_{q\Phi}$ and $T_{qP}$ are characteristic temperatures [29]. As it was shown recently [30] similar Barrett-type expressions can be used for AFM coefficient $a_i^L(T)$ of pure bismuth ferrite $a_i^L(T) = \alpha_T^{(L)} T_L \left( \coth\left( T_L/T \right) - \coth\left( T_L/T_N \right) \right)$ with the Neel temperature $T_N = 645$ K and characteristic temperature $T_L = 550$ K. The expression $L \sim \sqrt{a_1^L(T)/a_{11}}$, valid in the isotropic approximation, describes quantitatively both the temperature dependence of the AFM order parameter measured experimentally in BiFeO$_3$ by neutron scattering by Fischer et al. [26] and anomalous AFM contribution to the specific heat behaviour near the Neel temperature measured experimentally by Kallaev et al. [31].

The AFD-FE-AFM coupling energy $\Delta G_{BQC}$ is a biquadratic form of the order parameters $L_i$, $P_i$ and $\Phi_i$ (see Suppl. Mat in Ref. [27]):

$$\Delta G_{BQC} = \zeta_{ijkl} \Phi_i \Phi_j P_k P_l + \kappa_{ij} \Phi_i^2 L_j^2 + \lambda_{ij} P_i^2 L_j^2, \qquad (5)$$

For a given symmetry the coupling energy in Eq. (2d) includes unknown tensorial coefficients in Voight notations for the AFD-FE ($\zeta_{44}$, $\zeta_{11}$, $\zeta_{12}$) biquadratic couplings. Below, due to the lack of experimental data, FE-AFM and AFD-AFM rotomagnetic and biquadratic magnetoelectric coupling constants are assumed to be isotropic, $\lambda_{ij} = \lambda \delta_{ij}$ and $\kappa_{ij} = \kappa \delta_{ij}$.

The ELS energy in the *R3c* phase is

$$\Delta G_{ELS} = -\left( s_{ijkl} \sigma_{ij} \sigma_{kl} + Q_{ijkl} \sigma_{ij} P_k P_l + R_{ijkl} \sigma_{ij} \Phi_k \Phi_l + Z_{ijkl} \sigma_{ij} L_k L_l \right), \qquad (6)$$

Here $s_{ijkl}$ are elastic compliances tensor components, $Q_{ijkl}$ are electrostriction tensor components, $R_{ijkl}$ are rotostriction tensor components and $Z_{ijkl}$ are magnetostriction tensor components for AFM order parameter **L**. All coefficients in the thermodynamic potential (1)-(5) were extracted from experimental results in Ref.[27], except for rotostriction, electrostriction and magnetostriction ones, which were determined in this work from independent experimental data in **Appendix A.** So that all parameters included in the LGD functional parts (1)-(6) are extracted from experiments and not chosen within "a reasonable range".

Let us apply the thermodynamic approach based on the free energy (1)-(6) to a dense fine-grained BFO ceramics, for which the grain size $R$ varies from several tens nanometers to several microns, and the grains are separated by a stressed inter-grain shell of thickness $R_0 = (5 - 50)$ nm. [see **Fig.1(a)**]. The stresses can originate from many different sources, such as surface tension itself, as well as from chemical pressure in the regions enriched by e.g. oxygen vacancies and/or other defects such as Fe clusters**.** Below we will show that the contributions of both these sources into the total stress are additive, and, therefore hardly separable in many cases. This statement will be approved below mathematically.

For the case of densely packed spherical grains of radius $R$, which equatorial cross-section is shown in **Fig.1(b)**, the ratio of the grains volume to the inter-grain space can be elementary calculated as $\frac{6}{\pi} - 1 \approx 0.91$. The ratio $\eta_S$ of core volume to the total "shell + inter-grain" volume is smaller than 0.91, namely $\eta_S = \frac{8R^3 - (4/3)\pi(R - R_0)^3}{(4/3)\pi(R - R_0)^3} = \frac{6R^3}{\pi(R - R_0)^3} - 1$. Hence the significant part of the ceramics with densely packed identical spherical grains should be regarded affected by the surface, as well as by the chemical pressure created by the elastic defects accumulated in the grain shells and inter-grain spaces. In reality the grains are non-spherical, different in size and so packed much more densely reducing the part of the inter-grain space dramatically.

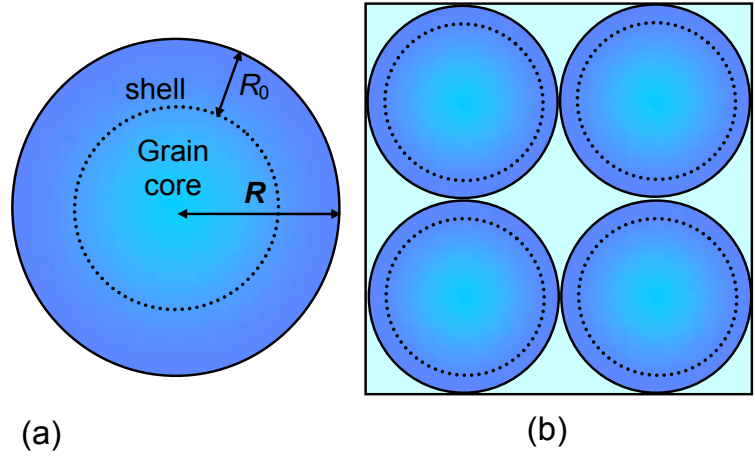

**FIGURE 1. (a)** Schematics of the spherical grain with radius $R$ covered by the shell of thickness $R_0$, where the stresses are accumulated. **(b)** Cross-section of the densely packed 8 spherical grains of radius $R$ placed inside the cube with edge 4R.

Therefore, for the case of the strained fine-grained ceramics Eq.(1) becomes affected by electrostrictive, rotostrictive and magnetostrictive couplings according to Eqs.(5)-(6). Namely, a formal expression for the possible shift of AFM transition temperature (we are interested in) related with the stresses near the grain boundaries can be obtained from the expression $a_j^{(L)}(T) = 0$, where the approximate formulae, $a_j^{(L)} \approx a_T^{(L)}(T - T_N) - Z_{klij}\sigma_{kl} + \kappa_{ij}\Phi_i^2 + \lambda_{ij}P_i^2$ is valid in the vicinity of Neel temperature. The expression is formal because the surface and gradient effects [33, 34, 37, 38] can contribute to the average values and their mean squire deviation in a complex and a priory nontrivial way. The concrete form of the expression for $a_j^{(L)}$ depends on the physical-chemical state of the grain bulk and surface.

Let us limit our consideration by the most common intrinsic surface stresses [32, 33, 34,] coupled with Vegard strains (chemical pressure) [8, 35, 36] acting on both polarization **P**, tilt **Φ** and AFM order parameter **L** via the electrostriction, rotostriction and magnetostricton couplings, respectively. Also we regard that depolarizing field acting on ferroelectric polarization inside the grain is negligibly small due to the screening charges. Within these assumptions the radial component of the excess chemical pressure (denoted as $\sigma_{rr}^W(r)$) and intrinsic surface stress (denoted as $\sigma_{rr}^\mu(r)$) inside the core and shell regions acquires the form derived in **Appendix D** and Refs.[33, 34, 37, 38] and:

$$\sigma_{rr}^\mu(r) \approx \begin{cases} -\dfrac{\mu}{R}, & r < R - R_0, \\ -\mu\left(\dfrac{1}{R - R_0} - \dfrac{1}{R}\right), & R - R_0 < r < R. \end{cases} \tag{7a}$$

$$\sigma_{rr}^{W}(r) \approx \begin{cases} 0, & r < R - R_0, \\ -G\dfrac{1+\nu}{1-\nu}W^m \delta N_m \dfrac{2R^3 + r^3}{3r^3}, & R - R_0 < r < R. \end{cases} \qquad (7b)$$

Here $r$ is the distance from the grain centre. The components of intrinsic surface stress tensor are regarded diagonal hereinafter, i.e. $\mu_{kl} = \mu \delta_{kl}$ and $\mu$ is about $(1-10)$N/m. Poisson ratio is $\nu = -s_{12}/s_{11}$ for cubic m3m symmetry. $G = (c_{11} - c_{12})/2$ is the shear modulus, $c_{jl}$ are elastic stiffness's modulus of the material [about $(10^{11} - 10^{12})$Pa].

$W_{kl}^{m}$ is Vegard strain tensor of $m$-type defects hereinafter regarded diagonal, $W_{kl}^{m} = W^m \delta_{kl}$. For perovskites ABO$_3$ the Vegard strain tensor is often related with vacancies and its absolute value can be estimated as $|W| \propto (5-20)$ Å$^3$ for oxygen and cation vacancies [36]. Note that Vegard tensor $W_{ij}$ is usually diagonal for oxygen vacancies in perovskites, but not isotropic [36]. Notably, "compositional" Vegard strains $\delta u_{kl} = W_{kl}^m \delta N_m$ can reach percents for vacancies concentration variation the near the surface $\delta N_m \sim 10^{27}$ m$^{-3}$, that corresponds to approximately one defect per 10 unit cells for the typical cell concentrations in perovskites $\sim 1.5 \times 10^{28}$ m$^{-3}$. Despite the concentration is much higher than the defect concentration in a bulk [39], such values are typical for vacancies segregation near the surface due to the strong lowering of their formation energy at the surface [40, 41].

Let us average the total stress $\sigma_{rr}(r) = \sigma_{rr}^{W}(r) + \sigma_{rr}^{\mu}(r)$ in Eqs.(7) over the grain volume $V = \dfrac{4}{3}\pi R^3$ under the condition $R_0 << R$. Using calculations listed in **Appendix D** the averaging yields

$$\langle \sigma_{rr}(R) \rangle \approx \eta \frac{R_0}{R}, \qquad \eta = -\frac{\mu}{R_0} - 3G\frac{1+\nu}{1-\nu}W^m \delta N_m. \qquad (8)$$

As one can see from the explicit form of the "total stress" parameter $\eta$, its first term ($\sim\mu$) originates from the intrinsic surface stress, and the second term ($\sim W^m \delta N_m$) originates from the excess chemical pressure. Thus Eq.(8) proves that the of both chemical pressure and surface tension sources of the stresses contribute into the total stress additively, and, therefore hardly separable in many cases.

Assuming that the coupling between AFM and AFD, AFM and FE order parameters are weak, the decoupling approximation is valid with high accuracy, and so the renormalized AFM transition temperature for a quasi-spherical grain of radius $R$ covered by a thin shell of thickness $R_0$ acquires the form [see **Appendix B**]:

$$T_{AFM} \approx T_N - \frac{1}{\alpha_T^{(L)}}\left[ \frac{\kappa(2R_{12} + R_{11})}{|a_{11}^{(\Phi)}|} + \frac{\lambda(2Q_{12} + Q_{11})}{|a_{11}^{(P)}|} - (2Z_{12} + Z_{11}) \right]\eta\frac{R_0}{R}. \qquad (9)$$

The shift of $T_N$ in Eq.(9) contains three contributions, rotomagnetic [proportional to $\kappa(2R_{12} + R_{11})$], rotoelectric [proportional to $\lambda(2Q_{12} + Q_{11})$] and magnetostrictive [proportional to $(2Z_{12} + Z_{11})$] coupling with the total stress $\langle\sigma_{rr}(R)\rangle \sim \eta\dfrac{R_0}{R}$. Estimations of the contributions for the parameters from **Table I** give for the following coefficients of rotomagnetic, rotoelectric and magnetostriction contributions

$$\frac{1}{\alpha_T^{(L)}}\frac{\kappa(2R_{12} + R_{11})}{a_{11}^{(\Phi)}} \cong -1.53\times10^{-8}\frac{Km^3}{J}, \quad \frac{1}{\alpha_T^{(L)}}\frac{\lambda(2Q_{12} + Q_{11})}{a_{11}^{(P)}} \cong 2.19\times10^{-9}\frac{Km^3}{J},$$

$$\frac{2Z_{12} + Z_{11}}{\alpha_T^{(L)}} \cong -1.2\times10^{-8}\frac{Km^3}{J}. \tag{10}$$

Rotomagnetic and rotoelectric coupling contributions to the shift (9) are different in sign because the sum $2R_{12} + R_{11} = -2.18\times10^{18}$ m$^{-2}$ is negative, and sum $Q_{11} + 2Q_{12} = 0.0235$ m$^4$/C$^2$ appeared positive for BFO (see **Table I**). According to the estimates (10) the highest is the rotomagnetic contribution, magnetostrictive one is a bit smaller, and the rotoelectric contribution is about an order of magnitude smaller.

**Table I**. LG potential for BiFeO$_3$

| Parameter | SI units | Value for BiFeO$_3$ | Reference |
|---|---|---|---|
| $a_{11}^{(\Phi)}$ | J/m$^7$ | $-4.53\times10^{49} + 4.5\times10^{48}\times \coth(300/T)$ | [27] |
| $a_{11}^{(P)}$ | m$^5$J/(C$^4$) | $-1.35\times10^9$ | [27] |
| $a_{111}^{(\Phi)}$ | J/m$^9$ | $16.72\times10^{70} - 3.4\times10^{70}\times \coth(400/T)$ | [27] |
| $a_{111}^{(P)}$ | m$^9$J/(C$^6$) | $11.2\times10^9$ | [27] |
| $\kappa$ | J/(A$^2$m$^3$) | $7.4\times10^{17}$ | [27] |
| $\lambda$ | J m$^3$/(A$^2$C$^2$) | $3.8\times10^{-4}$ | [27] |
| $\alpha_{LT}$ | J/(A$^2$m K) | $3.02\times10^{-6}$ | see Appendix C |
| $T_N$ | K | 645 | Neel temperature |
| $\beta_L$ | J m/A$^4$ | $1.03\times10^{-14}$ | see Appendix C |
| $Q_{ij}$ | m$^4$/C$^2$ | $Q_{11} + 2Q_{12} = 0.0235$ | see Appendix A |
| $R_{ij}$ | m$^{-2}$ | $2R_{12} + R_{11} = -2.18\times10^{18}$ | see Appendix A |
| $Z_{ij}$ | m$^2$/A$^2$ | $Z_{11} + 2Z_{12} = 3.65\ 10^{-14}$ | see Appendix C |

| $c_{ij}$ | Pa | $c_{11} = 3.02 \times 10^{11}$; $c_{12} = 1.62 \times 10^{11}$; $c_{44} = 0.68 \times 10^{11}$ | [42] |
|---|---|---|---|

We show the dependence of the AFM transition temperature $T_{AFM}$ on the grain radius $R$ in **Fig. 2(a)** and analyze its rotoelectric, magnetostrictive and rotomagnetic contributions in **Fig. 2(b)**.

From **Fig.2(b)** the size-induced increase of the AFM temperature is caused by the rotomagnetic and magnetostrictive couplings. The rotoelectric coupling leads to the decrease AFM transition, and the shift is several times smaller than the increase caused by rotomagnetic coupling accordingly to the estimates (10).

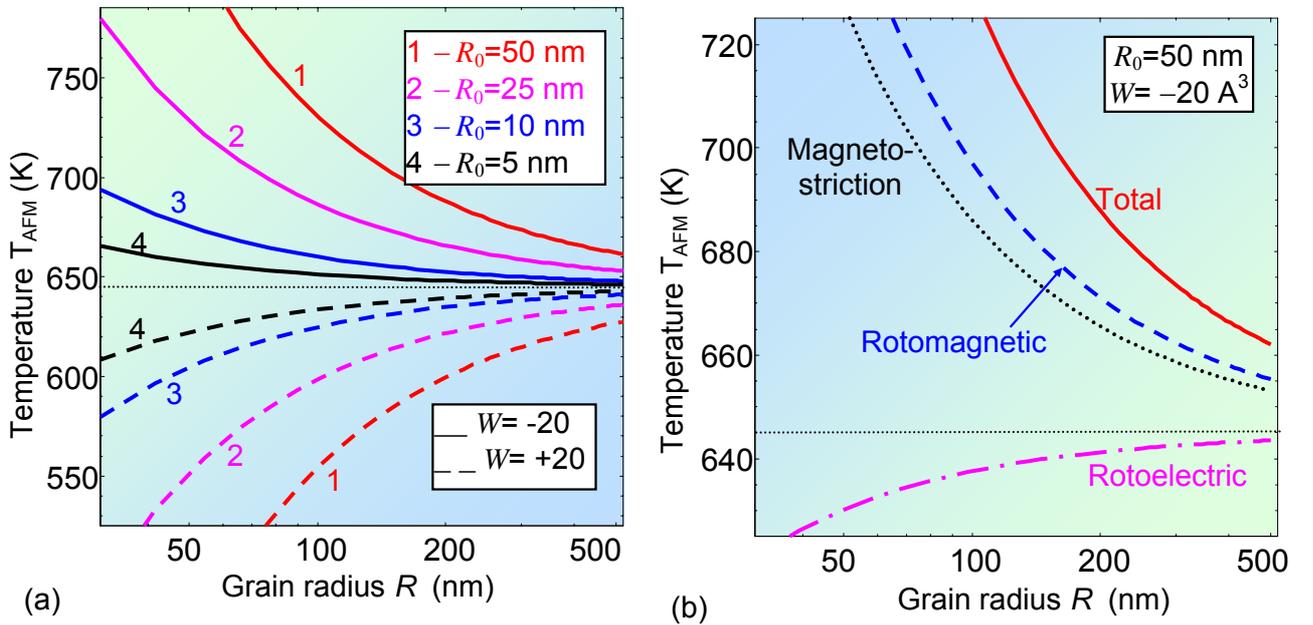

**FIGURE 2. (a)** Dependence of the AFM transition temperature $T_{AFM}$ vs. the grain radius R calculated from Eq.(8) for several shell thicknesses $R_0 = 50$ nm (curves 1), $R_0 = 25$ nm (curves 2), $R_0 = 10$ nm (curves 3), and $R_0 = 5$ nm (curves 4). Total Vegard coefficient $W = \sum_m W_m$ is equal to $-20$ Å³ for solid curves and $+20$ Å³ for dashed curves. **(b)** Separate contributions (rotomagnetic, magnetostrictive and rotoelectric) to the $T_{AFM}$. Surface tension coefficient $\mu = 5$ N/m and total defect concentration in the shell $\sum_m \delta N_m = 10^{27}$ m⁻³. Other parameters are taken from **Table 1**.

Color map of the AFM transition temperature $T_{AFM}$ in coordinates "grain radius R – shell thickness $R_0$" was calculated from Eq.(8) and shown in **Figs.3 (a, b, c)** for positive, zero and negative Vegard coefficient $W$, respectively. From the figures positive and zero $W$ decrease the transition temperature [see **Figs.3 (a, b)**], while only negative $W$ can increase it [see **Fig.3 (c)**]. The increase is significant for relatively small grain with radius less than 200 nm and thick shells with thickness more than 10 nm.

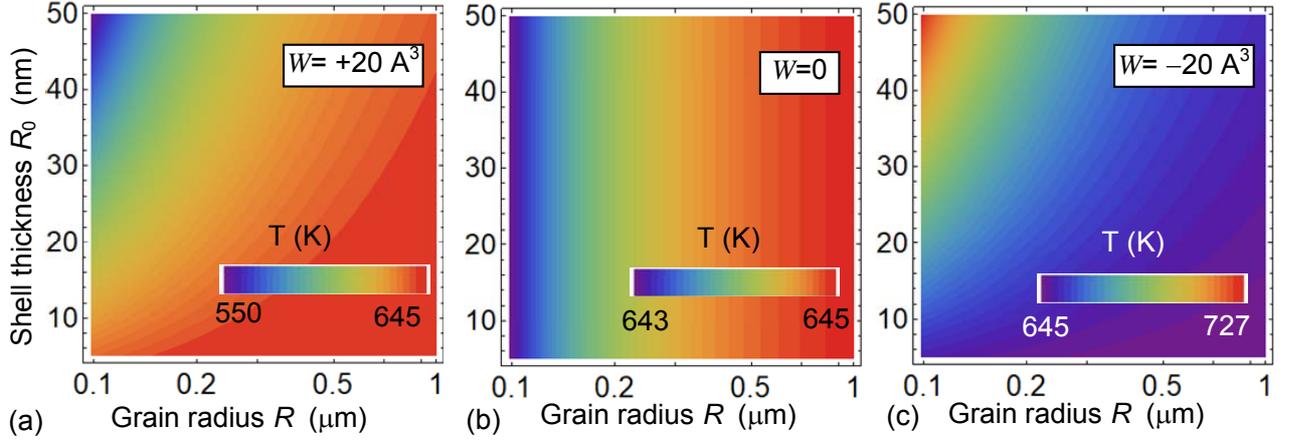

**FIGURE 3.** Color map of the AFM transition temperature $T_{AFM}$ in coordinates "grain radius R – shell thickness $R_0$" calculated from Eq.(9) for the same parameters as in **Fig.2** and **Table 1.**

### III. EXPERIMENTAL RESULTS

A polycrystalline BFO sample prepared by a two-stage solid-state reaction technique [43] was characterized by a single phase rhombohedral structure described by *R3c* space group [44]. The XRD data affirmed a chemical homogeneity of the compound with an accuracy of about 3% which is conditioned by the precision of a conventional X-ray diffractometer [**Fig. 4(c)**]. The synthesis conditions used to prepare the sample, viz. quite high final sintering temperature of 880°C applied for a short time period (10 min) allowed getting the high purity compound with a typical grain size of about 1-5 μm coexisting with an intergranular texture. The compound is characterized by an increased amount of the intergranular texture which volume fraction is about 1% as confirmed by the SEM measurements which is significantly larger that the values attributed to similar compounds prepared by conventional solid state reaction technique. It assumed that the structure of intergranular texture is highly defective because of a numerous dislocations, inhomogeneous stress distribution, local variations of the chemical composition while it's characterized by quite high chemical inhomogeneity as confirmed by the XRD measurements [**Fig. 4(c)**].

Temperature dependencies of magnetization were measured in zero field cooled (ZFC) and field cooled (FC cooling) modes in the temperature range 300 – 1000 K under magnetic field of 1kOe with a slow scan rate (2 seconds per measuring point, accuracy ~ 0.1 K). Small value of remnant magnetization [**Figs. 4(a,b)**] is associated with weak ferromagnetic state which becomes pronounced due to a disruption of the spatially modulated magnetic structure occurred in the vicinity of numerous structural defects specific for the compound. Temperature dependent magnetization measurements allowed to observe the significantly shifted antiferromagnetic transition temperature ($T_N$ ~690K) as compared to the widely noted value of 640 K specific for the single crystal BFO [45, 46] [see **Fig.4(b)**]. SEM images of the dense ceramics for different magnifications are shown in **Fig.4(d)-(f)**].

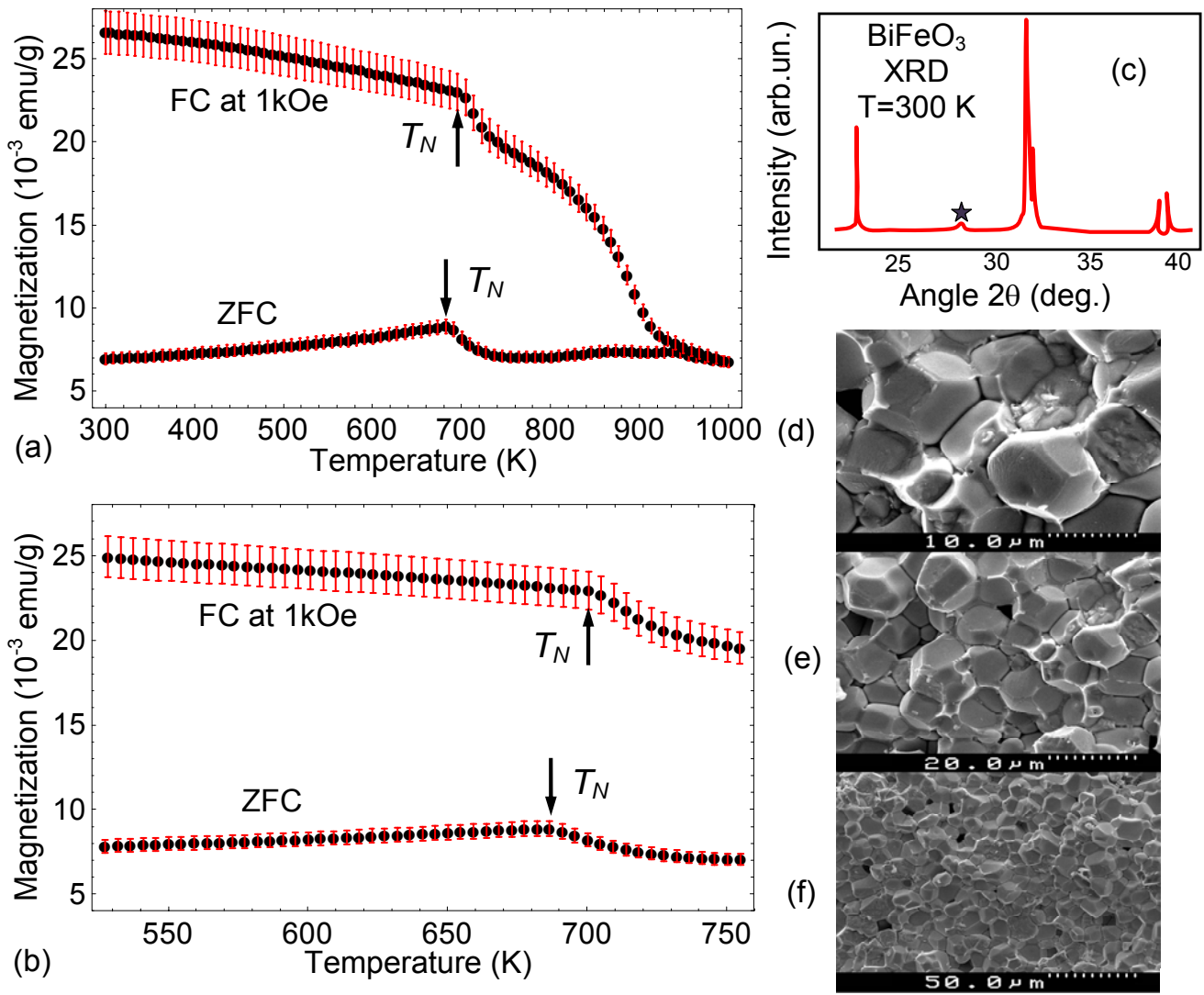

**FIGURE 4.** **(a,b)** Temperature dependencies of magnetization measured in BFO ceramics under FC condition (upper curve) and ZFC (lower curve) in a field of 1 kOe. Plot **(b)** is the zoomed region of the plot **(a)** in the vicinity of AFM temperature. Error bars are shown by red lines. XRD data shown in the inset **(c)** testify the phase purity of studied BFO ceramics (tiny amount of the impurity phase is marked by asterisk symbol). **(d)-(f)** SEM images of the dense ceramics for different magnifications.

It should be noted that the magnetic transition temperature is shifted towards high temperatures for both FC and ZFC curves, and the difference in the magnetic anomalies observed at both magnetization dependencies is about 10 K and cannot be caused by some drawbacks in the measuring procedure. FC and ZFC curves do not merge above $T_N$ because the magnetization data testify a presence of the magnetic impurity (viz. $\gamma$-$Fe_2O_3$ phase with a volume fraction of less than 1% as confirmed by the XRD measurements), which forms notable "background" into the magnetization curves lasting up to a temperature above 900 K, one should note similar behavior of the temperature dependencies of magnetization observed for the single-crystal and ceramic $BiFeO_3$ [45, 47]. Assuming negligibly small amount of the mentioned magnetic impurity one should not consider any significant effect on the magnetic transition temperature of the compound. The increased value of magnetic transition temperature can be explained by a joint action of magnetostriction and rotomagnetic effects which are usually very small in a homogeneous bulk BFO crystal and leading to the temperature shift of about $(0.5 - 5)$ K. These effects can be much more pronounced in ceramics due to the internal intergranular stresses. In this sense relevant phenomenology will allow some insight to the intrinsic stress and strain gradients.

## IV. DISCUSSION

To relate the above theoretical estimates with the experimental results shown in **Fig.4(a, b)** we assume that several types of defects (oxygen vacancies and Fe clusters) are accumulated in the shells and their influence is synergetic. Only for the case the total defect concentration in the shell can reach relatively high values, $\sum_m \delta N_m = 10^{27} m^{-3}$. In order to compare the above theory with the experimental results shown in **Fig.4(b)** the observable physical quantities (e.g. magnetization **M**) should be averaged over the grain radius $R$ and shell thicknesses $R_0$ with a definite normalized distribution function $f(R, R_0)$. Since the $M^2 \sim (T - T_{AFM})$ average transition temperature is given by expression:

$$\langle T_{AFM} \rangle = \int_{R_{min}}^{R_{max}} dR \int_{R_0^{min}}^{R_0^{max}} dR_0 f(R, R_0) T_{AFM}(R, R_0). \tag{11}$$

For instance, assuming that the shell thickness $R_0$ is constant, and the distribution of grain radius is quasi-homogeneous with $R_{min}$ and $R_{max}$, i.e. $\langle T_{AFM} \rangle = \int_{R_{min}}^{R_{max}} \frac{dR T_{AFM}(R, R_0)}{R_{max} - R_{min}}$, one obtains from Eq.(9)-(11) that

$$\langle T_{AFM} \rangle \approx T_N + \frac{1}{\alpha_T^{(L)}} \left[ \frac{\lambda(2Q_{12} + Q_{11})}{|a_{11}^{(P)}|} + \frac{\kappa(2R_{12} + R_{11})}{|a_{11}^{(\Phi)}|} - (2Z_{12} + Z_{11}) \right] \frac{\eta}{2\Delta R} \ln\left( \frac{\langle R \rangle + \Delta R}{\langle R \rangle - \Delta R} \right) \tag{12}$$

Where $\langle R \rangle = \frac{R_{max} + R_{min}}{2}$, $\Delta R = \frac{R_{max} - R_{min}}{2}$ and $\Delta R = \langle R \rangle - R_{min}$. From **Fig. 5** the dependence of the averaged AFM transition temperature $\langle T_{AFM} \rangle$ on $R_{max}$ is shown for $R_0 \approx 45$ nm, $R_{min} \approx 50$ nm and $W = -20$ Å$^3$. The values $R_0$ and $R_{min}$ where taken for illustration, they are within reasonable ranges $5 \leq R_0 \leq 50$ nm [38**Ошибка! Закладка не определена.**] and $50 < R_{\min} < 500$ nm typical for sub-micro and nanograined ceramics and satisfy the necessary condition $R_0 \leq R_{\min}$. According to **Fig. 5** the increase of $\langle T_{AFM} \rangle$ above 45 K is possible for the ceramic with the average grain radius below 150 nm. However according the **Fig. 5** for the ceramics with the average grain size about 5 μm the Neel temperature should be about 650 K that is close to the single crystal value 645 K.

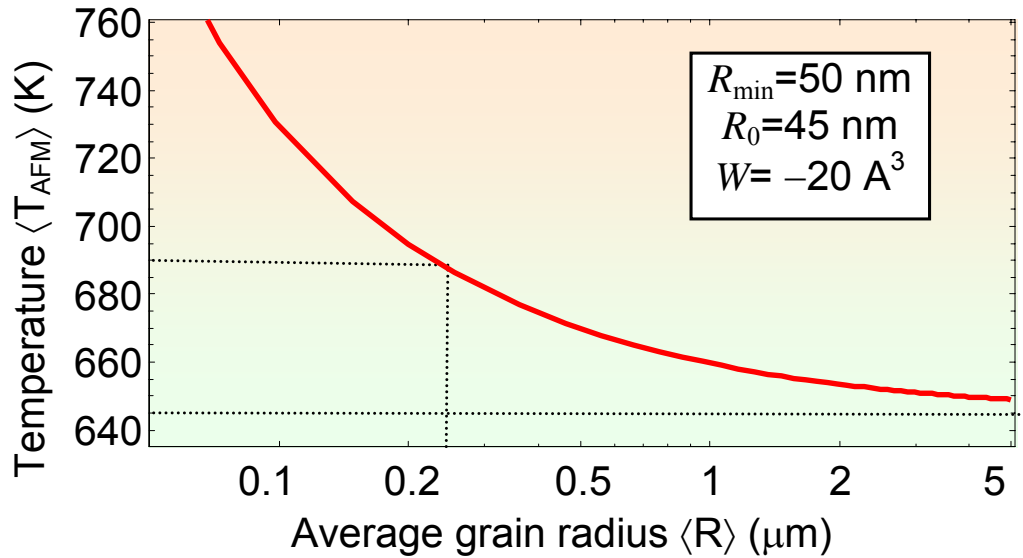

**FIGURE 5.** The dependence of the averaged transition temperature $\langle T_{AFM} \rangle$ on the average grain radius $\langle R \rangle$ calculated from Eq.(10) for minimal grain radius $R_{\min} \approx 50$ nm, shell thickness $R_0 \approx 45$ nm, and Vegard coefficient $W = -20$ Å$^3$. Other parameters are the same parameters as in **Fig.2** and **Table 1**.

Hence the proposed theoretical model can explain experimental data shown in **Fig.4(a)** only qualitatively, because it gives the increase of $\langle T_{AFM} \rangle$ above 45 K for fine-grained ceramics with significant amount of grains with radius smaller than 250 nm. The one order of magnitude discrepancy between the average grain sizes required from the theoretical model (less than 500 nm) and experiment (about 5 μm) to reach increase of $\langle T_{AFM} \rangle$ above 45 K evidently speaks in favor of strongly underestimated impact of the rotomagnetic coupling by the model parameters or unexpectedly high contribution of the small grains into the average magnetization (non-uniform distribution function of the grain sizes).

## V. SUMMARY

Using Landau-Ginzburg-Devonshire theory for BiFeO$_3$ dense ceramics with quasi-spherical micron sized grain cores and nanosized inter-grain spaces we calculated a surprisingly strong size-induced increase of the AFM transition temperature caused by the joint action of rotomagnetic effect and magnetostriction coupled with elastic stresses accumulated in the inter-grain spaces. The rotoelectric coupling leads to the decrease of AFM transition temperature, and the shift is several times smaller than the increase caused by rotomagnetic coupling.

Also we performed experiments for dense $BiFeO_3$ ceramics, which revealed that the AFM transition was observed at $T_N \sim 690$ K instead of $T_N \sim 645$ K for a single crystal. To explain qualitatively the result we consider the possibility to control AFM properties of multiferroic $BiFeO_3$ via biquadratic antiferrodistortive rotomagnetic, rotoelectric and magnetoelectric couplings. To reach quantitative agreement between the theoretical model and experimental data one could also consider low symmetry phases [48, 49] with possibly higher impact of the rotomagnetic coupling and other LG parameters.

**Acknowledgments.** This project (A.N.M., O.M.F. and D.V.K.) has received funding from the European Union's Horizon 2020 research and innovation programme under the Marie Skłodowska-Curie grant agreement No 778070 – TransFerr – H2020-MSCA-RISE-2017. A.N.M. work was partially supported by the National Academy of Sciences of Ukraine (projects No. 0117U002612 and No. 0118U003375). V.G. acknowledges the National Science Foundation grant number DMR-1420620. D.V.K. and M.V.S. acknowledge MK-1720.2017.8 and RFFI (# 17-58-45026)

# SUPPLEMENT

## Appendix A. Rotostriction and electrostriction coefficients for BiFeO₃

**Orthorhombic and "cubic" phases.** Let us consider antiferro-distortive phase of BiFeO3 with orthorhombic symmetry (Pbnm space group). This unipolar phase is characterized by the anti-phase tilts of octahedral groups of oxygen ions in all three independent directions. Lattice constant temperature dependences from Arnold et al [50] is shown in **Fig.S1**.

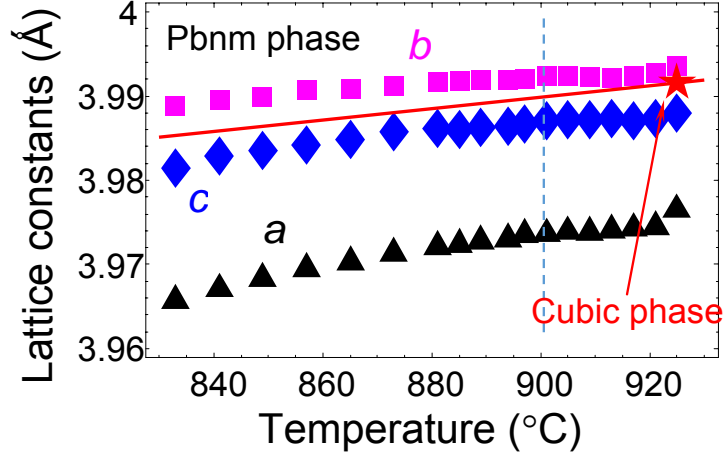

**Fig. S1.** Temperature dependences of the pseudo-cubic lattice constant from Arnold et al. [50]. Three different lattice constants for orthorhombic phase is shown, namely $a$ (triangles), $b$ (boxes) and $c$ (diamonds). Star represents the lattice constant $a_0$ of hypothetical cubic phase after Palai et al. [51], while solid line is for the extrapolation of $a_0$ to lower temperature (see text). Note that the exact structure of the phase above 900 °C is not known for certain.

In order to estimate the spontaneous strain components one should use an extrapolated lattice constant of high temperature "virtual" cubic phase in the following form

$$a_0 \approx a_C \left(1 + \alpha_T (T - 925)\right) \tag{S.1}$$

Since available experimental data (see **Fig. S1**) do not allow determination of thermal expansion $\alpha_T$ for cubic phase, we'll use $\alpha_T$ as a fitting parameter. The spontaneous strain components could be found from lattice constants as follows:

$$u_{11} = \frac{a}{a_0} - 1, \quad u_{22} = \frac{b}{a_0} - 1, \quad u_{33} = \frac{c}{a_0} - 1 \tag{S.2}$$

The temperature dependences of strain components are presented in Fig. S2.

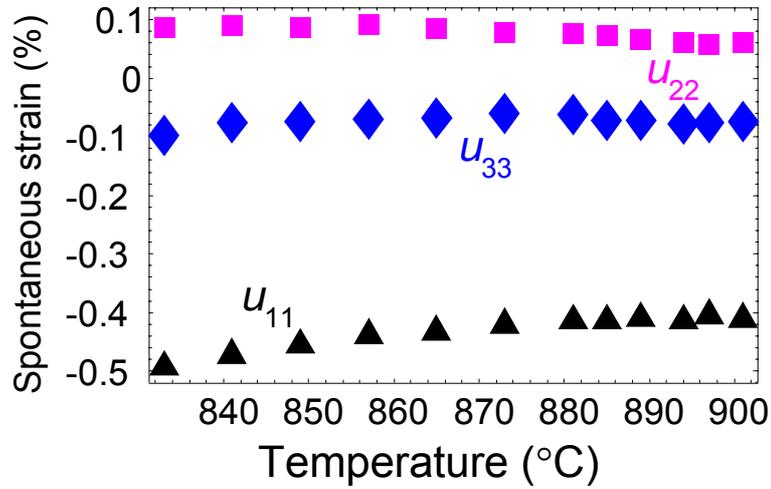

**Fig. S2.** Temperature dependences of the spontaneous strain components recalculated from the data presented in **Fig. S1.**

In order to get the rotostriction coefficients from the experimental results (Fig.S2) one has to consider temperature dependences of order parameters of Pbnm phase (see Fig. S3). Here we express tilt vector components via corresponding displacements of oxygen ions.

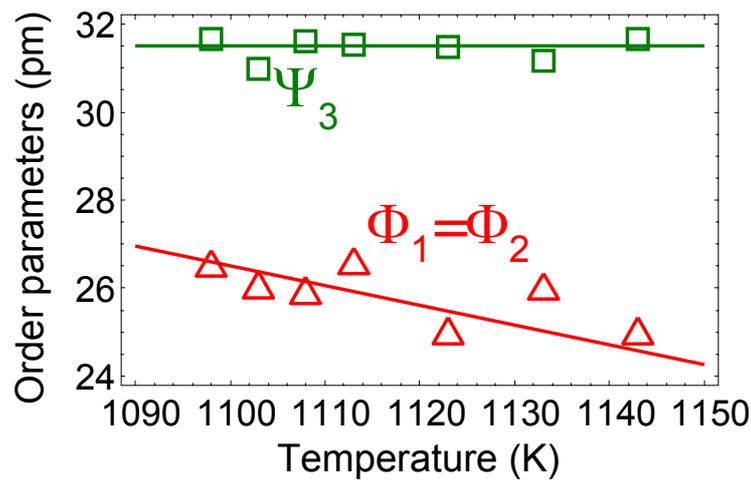

**Fig. S3.** Temperature dependences of order parameters for orthorhombic phase from Arnold et al [50], namely out-of phase tilts (triangles) and in-phase tilts (boxes). Lines represents simple fitting with linear temperature dependences.

Here we fit "tilts" with simple linear temperature dependence since tilt variation is small in all the temperature region of interest. Next we recall the phenomenological relations between the lattice constants (spontaneous strain) and "tilts". Here we used so called orthorhombic setting corresponds to the rotation of the psedo-cubic unit cell on the angle $\pi/4$ around rseudocubic axis [001], so that unit cell is the rectangular parallelepiped with sizes a, b and c.

$$a = a_0\left(1 + \left(R_{11} + R_{12} - \frac{R_{44}}{2}\right)\Phi_1^2 + K_{12}\Psi_3^2\right), \qquad \text{(S.3a)}$$

$$b = a_0\left(1 + \left(R_{11} + R_{12} + \frac{R_{44}}{2}\right)\Phi_1^2 + K_{12}\Psi_3^2\right), \qquad \text{(S.3b)}$$

$$c = a_0\left(1 + 2R_{12}\Phi_1^2 + K_{11}\Psi_3^2\right). \qquad \text{(S.3c)}$$

In addition, it is useful to introduce strain in pseudo-cubic reference frame:

$$u_{ab} = \frac{b - a}{a_0} = R_{44}\Phi_1^2, \qquad \text{(S.4a)}$$

$$u_{aa} = \frac{b + a}{2a_0} - 1 = \left(R_{11} + R_{12}\right)\Phi_1^2 + K_{12}\Psi_3^2, \qquad \text{(S.4b)}$$

$$u_{33} = \frac{c}{a_0} - 1 = 2R_{12}\Phi_1^2 + K_{11}\Psi_3^2. \qquad \text{(S.4c)}$$

The results of the fitting with Eqs. (S.4) are presented in Figs.S4, the obtained phenomenological constants of rotostriction coupling are summarized in Table S1.

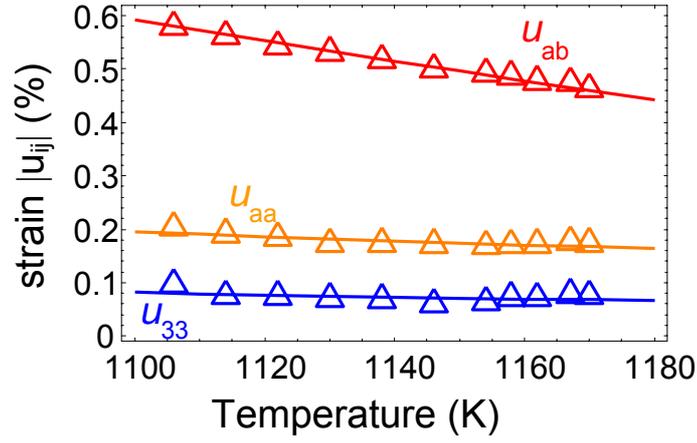

**Fig. S4.** Temperature dependences of different components of spontaneous strain (triangles) along with the fitting (lines) on the base of phenomenological relation (S.4). Only absolute values of strains are shown for clarity.

**Table S1.** Rotostriction coefficients in $10^{18}$ m$^{-2}$.

| Coefficient | $R_{11}$ | $R_{12}$ | $R_{44}$ | $K_{11}$ | $K_{12}$ |
|---|---|---|---|---|---|
| value | −1.32 | −0.43 | 8.45 | −0.21 | −0.72 |

**Rhombohedral phase R3c.** Rhombohedral phase of BiFeO$_3$ is polar and could be characterized with two order parameters, polarization and out-of-phase tilts directed along [111] direction of initial lattice of aristophase (hypothetical cubic phase). Therefore, order parameters components

$$\Phi_1^2 = \Phi_2^2 = \Phi_3^2 = \Phi_S^2/3 \tag{S.5a}$$

$$P_1^2 = P_2^2 = P_3^2 = P_S^2/3 \tag{S.5b}$$

Here $\Phi_S$ and $P_S$ are spontaneous tilt and polarization respectively. Lattice constant of R3c phase is

$$a_R = a_0\left(1 + \frac{R_{11} + 2R_{12}}{3}\Phi_S^2 + \frac{Q_{11} + 2Q_{12}}{3}P_S^2\right) \tag{S.6}$$

Using available experimental results (see **Fig.S5a**) for lattice constant and temperature dependences of spontaneous tilt and polarization obtained earlier (see e.g. [Karpinskii et al. NPG]), we fitted spontaneous strain of R3c phase on the basis of Eq.(S.6) (see **Fig. S5b**)

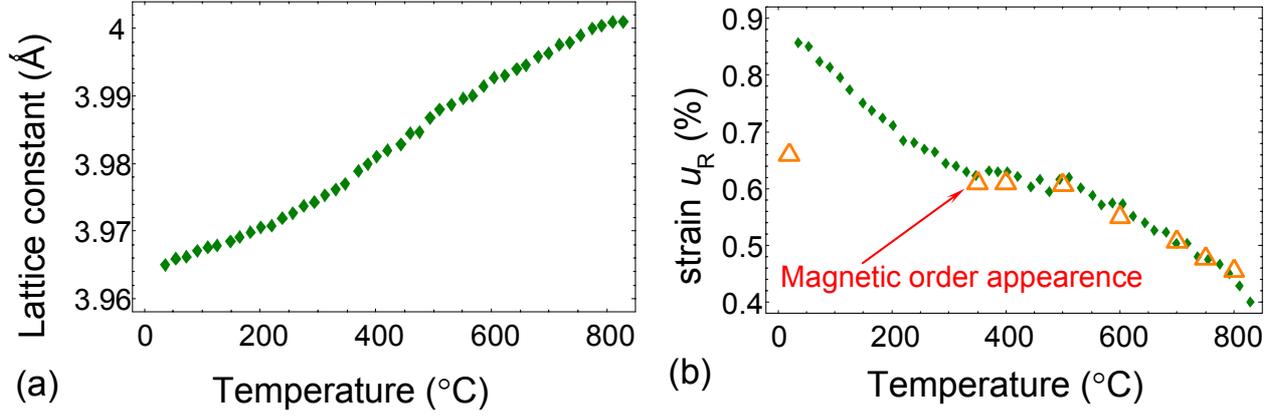

**Fig. S5.** Temperature dependences of (a) lattice constant of R3c phase of BiFeO3 (from [51]); (b) spontaneous strain, obtained from experimental data (diamonds) along with the fitting on the base of phenomenological relation (S.6) with parameter $Q_{11} + 2Q_{12} = 0.0235$ m$^4$/C$^2$ (triangles).

The deviation of calculated strain (triangles) from experimentally derived one (diamonds) below about 350 °C could be explained by the emergence of antiferromagnetic phase.

## Appendix B. Derivation of transition temperature

Using the scalar approximation for all tensors in Eqs.(2)-(7) and expression (8) for intrinsic stresses, the order-parameter dependent part of the free energy (1) acquires the form

$$\Delta G = \begin{pmatrix} \left[\alpha_{\Phi T}\left(T - T_{\Phi}\right) - 2\sigma_{rr}\left(2R_{12} + R_{11}\right)\right]\dfrac{\Phi^2}{2} + \beta_{\Phi}\dfrac{\Phi^4}{4} + \gamma_{\Phi}\dfrac{\Phi^6}{4} + \\ \left[\alpha_{TP}\left(T - T_C\right) - 2\sigma_{rr}\left(2Q_{12} + Q_{11}\right)\right]\dfrac{P^2}{2} + \beta_P\dfrac{P^4}{4} + \gamma_P\dfrac{P^6}{4} + \\ \left[\alpha_{LT}\left(T - T_{N0}\right) - 2\sigma_{rr}\left(2Z_{12} + Z_{11}\right)\right]\dfrac{L^2}{2} + \beta_L\dfrac{L^4}{4} + \gamma_L\dfrac{L^6}{4} + \\ + \zeta\Phi^2 P^2 + \left(\kappa\Phi^2 + \lambda P^2\right)L^2 \end{pmatrix} \tag{S.7}$$

Here the coefficients $\alpha_{LT} = 2\alpha_T^{(L)}$, $\alpha_{TP} = 2\alpha_T^{(P)}$, $\alpha_{\Phi T} = 2\alpha_T^{(\Phi)}$, $\beta_{\Phi} = 2a_{11}^{(\Phi)}$, $\beta_P = 2a_{11}^{(P)}$, etc.

Assuming that the coupling between AFM and AFD, AFM and FE order parameters are weak, the decoupling approximation is valid in Eq.(9) with high accuracy,

$$\left[\alpha_{\Phi T}\left(T - T_{\Phi}\right) - 2\sigma_{rr}\left(2R_{12} + R_{11}\right)\right]\Phi + \beta_{\Phi}\Phi^3 + \gamma_{\Phi}\Phi^5 = 0, \tag{S.8a}$$

$$\left[\alpha_{TP}\left(T - T_C\right) - 2\sigma_{rr}\left(2Q_{12} + Q_{11}\right)\right]P + \beta_P P^3 + \gamma_P P^5 = 0, \tag{S.8b}$$

From these equations the approximate expressions for the spontaneous values are

$$\Phi_S^2 = \frac{-\beta_\Phi - \sqrt{\beta_\Phi^2 - 4\gamma_\Phi \left(\alpha_{\Phi T}\left(T - T_\Phi\right) - 2\sigma_{rr}\left(2R_{12} + R_{11}\right)\right)}}{2\gamma_\Phi} \approx \Phi_0^2 - \frac{2\sigma_{rr}\left(2R_{12} + R_{11}\right)}{\left|\beta_\Phi\right|} \qquad \text{(S.9a)}$$

$$P_S^2 = \frac{-\beta_P - \sqrt{\beta_P^2 - 4\gamma_P \left(\alpha_{PT}\left(T - T_C\right) - 2\sigma_{rr}\left(2Q_{12} + Q_{11}\right)\right)}}{2\gamma_P} \approx P_0^2 - \frac{2\sigma_{rr}\left(2Q_{12} + Q_{11}\right)}{\left|\beta_P\right|}. \qquad \text{(S.9b)}$$

where $\Phi_0^2 = \dfrac{-\beta_\Phi - \sqrt{\beta_\Phi^2 - 4\gamma_\Phi \alpha_{\Phi T}\left(T - T_\Phi\right)}}{2\gamma_\Phi}$ and $P_0^2 = \dfrac{-\beta_P - \sqrt{\beta_P^2 - 4\gamma_P \alpha_{PT}\left(T - T_C\right)}}{2\gamma_P}$. So the

renormalized coefficient for AFM order and AFM transition temperature acquires the form:

$$a_R^{(L)} \approx \alpha_{LT}\left(T - T_N\right) - 2\sigma_{rr}\left[\left(2Z_{12} + Z_{11}\right) - \left(\frac{\lambda\left(2Q_{12} + Q_{11}\right)}{\left|\beta_P\right|} + \frac{\kappa\left(2R_{12} + R_{11}\right)}{\left|\beta_\Phi\right|}\right)\right], \qquad \text{(S.10)}$$

In the set of coefficients used in the main text

$$a_R^{(L)} \approx 2\alpha_T^{(L)}\left(T - T_N\right) - 2\sigma_{rr}\left[\left(2Z_{12} + Z_{11}\right) - 2\left(\frac{\lambda\left(2Q_{12} + Q_{11}\right)}{2\left|a_{11}^{(P)}\right|} + \frac{\kappa\left(2R_{12} + R_{11}\right)}{2\left|a_{11}^{(\Phi)}\right|}\right)\right], \qquad \text{(S.10b)}$$

Where we already used that $T_N = T_{N0} + \dfrac{\kappa\Phi_0^2 + \lambda P_0^2}{\alpha_{LT}} \approx 645$ K.

## Appendix C. Antiferromagnetic phase in BiFeO$_3$

In the antiferromagnetic phase additional order parameter, (anti)magnetization $L_s$ appears. Here we consider free energy dependence on the order parameter $L$:

$$\Delta G_L = \frac{\alpha_L}{2}L^2 + \frac{\beta_L}{4}L^4 - Z\sigma L^2 \qquad \text{(C.1)}$$

Here $\alpha_L$ is temperature dependent, while $\beta_L$ is usually constant, $\sigma$ is the elastic stress tensor, $Z$ is the antimagneto-striction coefficient. Spontaneous value of the (anti)magnetization is

$$L_S = \sqrt{-\alpha_L/\beta_L}. \qquad \text{(C.2)}$$

Corresponding contribution to free energy for mechanically free system is

$$\Delta G_L\big|_{\sigma=0} = \frac{-\alpha_L^2}{4\beta_L}. \qquad \text{(C.3)}$$

Contribution to the specific heat could be calculated as $\Delta C_L = -T\,\partial^2 \Delta G_L / \partial T^2$. We considered linear dependence on temperature of expansion coefficient

$$\alpha_L = \alpha_{LT}\left(T - T_N\right) \qquad \text{(C.4)}$$

Hence, we could derive from (C.3) and (C.4) the following relation:

$$\Delta C_L \big|_{T=T_N} = T_N \frac{(\alpha_{LT})^2}{2\beta_L} \qquad (C.5)$$

Next we'll use estimations of magnetic moment

$$L_0 \propto \frac{g \, s \, \mu_B}{V_{cell}} \qquad (C.6a)$$

Here $g$ is Landé g-factor, $s$ is the spin moment, $\mu_B = 9.274 \times 10^{-24}$ J/T is the Bohr magneton and $V_{cell}$ is the volume of elementary unit cell. At the same time, as it follows from Eq.(C.2)

$$L_0 = \sqrt{\frac{\alpha_{LT} T_N}{\beta_L}} \propto \frac{g \, s \, \mu_B}{V_{cell}} \qquad (C.6b)$$

Using $g=2$, $s=3/2$ and $V_{cell} =64$ Å$^3$ we obtained $L_0 =4.35 \ 10^5$ A/m. At the same time, one could estimate the drop at the Neel point $\Delta C_L \big|_{T=T_N} \approx 11$ J/(mol K) from experimentally measured temperature dependence of heat capacity (see e.g. [52], [53], while Ref.[54] give smaller value) . Then using Eqs.(C.5) and (C.6b) one could estimate coefficients $\alpha_{LT}$ and $\beta_L$ (see Table I in the main text).

The antiferromagnetic order appearance is accompanied by the lattice distortion. Comparing phenomenological relation for lattice constant of R3c phase

$$a_R = a_0 \left( 1 + \frac{R_{11} + 2R_{12}}{3} \Phi_S^2 + \frac{Q_{11} + 2Q_{12}}{3} P_S^2 + \frac{Z_{11} + 2Z_{12}}{3} L_S^2 \right) \qquad (C.7)$$

with experimentally observed dependence (see Fig.S5a) one could get the value of "hydrostatic" trace of antiferromagnetic striction tensor $Z_{11} + 2Z_{12}$ (see Fig.S6).

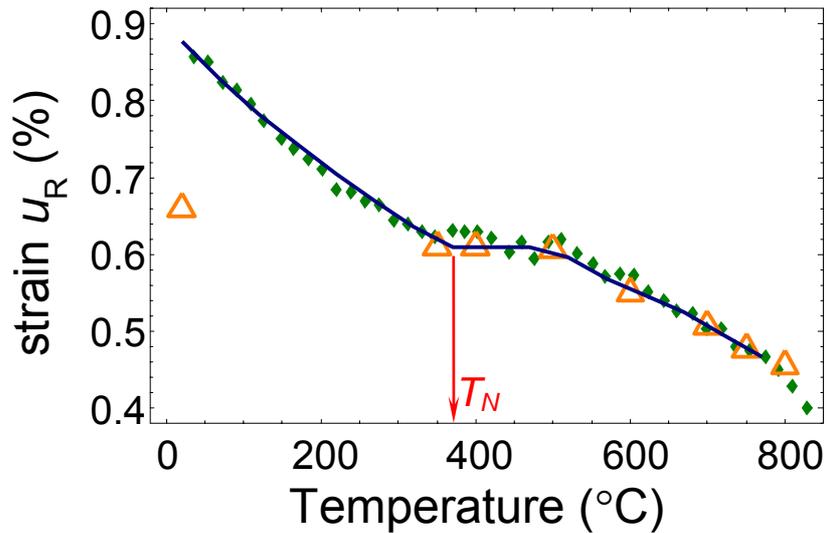

**Fig. S6.** Temperature dependences of spontaneous strain of R3c phase of BiFeO3 (from [51]) obtained from experimental data (diamonds) along with the fitting on the base of phenomenological relation (S.6) with parameter $Q_{11} + 2Q_{12} = 0.0235$ m$^4$/C$^2$ for two case $Z_{11} + 2Z_{12} = 0$ (triangles) and $Z_{11} + 2Z_{12} = 3.65 \cdot 10^{-14}$ (m/A)$^2$ (solid curve).

## Appendix D. Spherical core covered by a strained defect shell

Dilatation centre with equal distortion, $W_{ij} = W\delta_{ij}$, can be considered as a simple elastic model of impurity atom or vacancy whose own distortion (Vegard strain) is [55, 56, 57]

$$w_{xx}^k(\mathbf{r}) = w_{yy}^k(\mathbf{r}) = w_{zz}^k(\mathbf{r}) = W\delta(\mathbf{r} - \mathbf{r}_k) \qquad (D.1)$$

In fact $W$ is the volume change in the point of defect localization $\mathbf{r} = \mathbf{r}_k$. Nonzero components of the elastic displacement, strain and stressess induced by a spherically-symmetric elastic point defect (e.g. dilatation centre) located in the coordinate origin, $\mathbf{r} = 0$, have the form [56, 57]

$$\vec{u}^W(\mathbf{r}) = -\frac{(1+\nu)W}{4\pi(1-\nu)}\frac{\mathbf{r}}{r^3} \equiv \frac{(1+\nu)W}{4\pi(1-\nu)}\nabla\left(\frac{1}{r}\right), \qquad (r = \sqrt{x_1^2 + x_2^2 + x_3^2} \neq 0) \qquad (D.2a)$$

$$u_{ii}^W(\mathbf{r}) = -\frac{(1+\nu)W}{4\pi(1-\nu)}\frac{\partial}{\partial x_i}\left(\frac{x_i}{r^3}\right) = -\frac{(1+\nu)W}{4\pi(1-\nu)}\frac{r^2 - 3x_i^2}{r^5} \equiv \frac{(1+\nu)W}{4\pi(1-\nu)}\frac{\partial^2}{\partial x_i^2}\left(\frac{1}{r}\right), \qquad (D.2b)$$

$$u_{ij}^W(\mathbf{r}) = -\frac{(1+\nu)W}{4\pi(1-\nu)}\frac{\partial}{\partial x_i}\left(\frac{x_j}{r^3}\right) = \frac{(1+\nu)W}{4\pi(1-\nu)}\frac{3x_i x_j}{r^5} \qquad (i \neq j) \qquad (D.2b)$$

$$\sigma_{ii}^W(\mathbf{r}) = -\frac{G(1+\nu)W}{2\pi(1-\nu)}\frac{r^2 - 3x_i^2}{r^5} \equiv \frac{G(1+\nu)W}{4\pi(1-\nu)}\frac{\partial^2}{\partial x_i^2}\left(\frac{1}{r}\right), \qquad (D.2c)$$

$$\sigma_{ij}^W(\mathbf{r}) = -\frac{G(1+\nu)W}{2\pi(1-\nu)}\frac{3x_i x_j}{r^5} \equiv -\frac{G(1+\nu)W}{4\pi(1-\nu)}\frac{\partial}{\partial x_i}\left(\frac{x_j}{r^3}\right) \quad (i \neq j). \qquad (D.2d)$$

The Poisson ratio is $\nu = -s_{12}/s_{11}$ for cubic m3m symmetry. $G$ is the shear modulus. Radial stresses could be obtained as follows

$$\sigma_{rr}^W(r) = G\frac{(1+\nu)W}{(1-\nu)}N_d \begin{cases} 0, & r < R; \\ \dfrac{-2R^3 - r^3}{3r^3}, & R < r < R + R_0; \\ -2\dfrac{R^3 - (R+R_0)^3}{3r^3}, & r > R + R_0. \end{cases} \qquad (D.3)$$

When deriving Eq.(D.3) we used the integral

$$\int_0^{2\pi} \frac{d\varphi}{4\pi\sqrt{\left(\tilde{r}\cos\theta - r\right)^2 + \tilde{r}^2\sin^2\theta}} \int_0^{\pi}\sin\theta d\theta \int_R^{R+R_0}\tilde{r}^2 d\tilde{r} = \int_R^{R+R_0}\tilde{r}^2 d\tilde{r}\int_0^{\pi}\frac{\sin\theta d\theta}{2\sqrt{\left(\tilde{r}\cos\theta - r\right)^2 + \tilde{r}^2\sin^2\theta}} =$$

$$= \int_R^{R+R_0}\frac{\left(\tilde{r}+r\right) - \left|\tilde{r}-r\right|}{2r}\tilde{r}d\tilde{r} = \begin{cases} \dfrac{\left(R+R_0\right)^2 - R^2}{2}, & r < R; \\[2mm] \dfrac{r^3 - R^3}{3r} + \dfrac{\left(R+R_0\right)^2 - r^2}{2}, & R < r < R+R_0; \\[2mm] \dfrac{\left(R+R_0\right)^3 - R^3}{3r}, & r > R+R_0. \end{cases}$$ (D.4)

Next let us calculated the elastic fields created by the defects in the grain core ($0 \le r \le R_0$), inside the shell ($R < r \le R+R_0$) and outside the shell ($r > R+R_0$). For the purpose one should average the solution (D.3) over the spherical shell.

Since $\quad \dfrac{1}{V}\int_R^{R+R_0}r^2 dr\int_0^{\pi}\sin\theta d\theta\int_0^{2\pi}d\varphi\left(\dfrac{2R^3+r^3}{3r^3}\right) = \dfrac{1}{V}\left(\dfrac{4\pi R^3}{3}2\ln\left(\dfrac{R+R_0}{R}\right) + \dfrac{4\pi R^2 R_0}{3}\right) \approx \dfrac{3R_0}{R}\quad$ at

$R_0 \ll R$, let us average the total stress $\sigma_{rr}(r) = \sigma_{rr}^W(r) + \sigma_{rr}^\mu(r)$ in Eqs.(7) over the grain volume $V = \dfrac{4}{3}\pi R^3$ under the condition $R_0 \ll R$.

$$\left\langle\sigma_{rr}(r)\right\rangle \approx \eta\frac{R_0}{R}, \qquad\qquad \eta = -\frac{\mu}{R_0} - 3G\frac{1+\nu}{1-\nu}W^m\delta N_m.$$ (D.5)

In the approximate equality in Eqs.(D.5) we used that $R_0 \ll R$ and so one can neglect the difference $\dfrac{1}{R-R_0} - \dfrac{1}{R} \approx \dfrac{R_0}{R^2}$. Along with the values $\mu = 5$ N/m, $\delta N = 10^{27}\,\text{m}^{-3}$ and $\left|W\right| = 20$ Å$^3$ the estimates of the total stress excess give the values $\left|\eta\right| \cong 1.3\times10^{10}$ Pa (or J/m$^3$).